\documentstyle[12pt]{article}

\newcommand{\be}{\begin{equation}}
\newcommand{\br}{\begin{eqnarray}}
\newcommand{\ee}{\end{equation}}
\newcommand{\er}{\end{eqnarray}}

\begin{document}
\title{On Factorizing Correlation Functions in String Theory Using
Loop Variables}
\author{B. Sathiapalan\\ {\em
Physics Department}\\{\em Penn State University}\\{\em 120
Ridge View Drive}\\{\em Dunmore, PA 18512}}
\maketitle
\begin{abstract}
Factorization of string amplitudes is one way of constructing string
 interaction
vertices. We show that correlation functions in string theory can be
conveniently factorized using loop variables representing delta
 functionals.
We illustrate this construction with some examples where one
particle is off-shell.
These vertices are ``correct'' in the sense that they are guaranteed, by
construction, to reproduce S-matrix elements when combined with
 propagators
in a well defined way.
\end{abstract}
\newpage
\section{Introduction}
  The problem of obtaining a space-time picture for string theory
seems to be an outstanding one.  Except in unrealistic situations
it seems difficult to find exact solutions even of the tree level
equations in string theory.  For general backgrounds, the only convenient
method seems to be the sigma-model renormalization group method
\cite{l,cc,as,ft,cz,ds,ao,p,bm,ac,do}.
These methods break down as soon as one probes the high energy behaviour
or ask questions about massive modes \cite{ds2}.  Recent developments
in the ``duality'' symmetries that string theory is expected to possess
\cite{as2,w} make it very clear that one has to deal with all the
modes of the string, more or less simultaneously, if one is to have
 any hope
of understanding the underlying picture.  These results are
making it even more obvious than before that a string is more
than an infinite collection of particles of different masses.  This
fact was already obvious from the old ``duality'', namely
s-channel - t-channel duality and also from related properties like
modular invariance.

In order to deal with all the modes on an equal footing one has to go
off-shell. One has to understand the gauge invariance
associated with these (massive) modes.  Furthermore it is important to be
able to do this without losing contact with the low energy $\sigma $-
model description - since it is only there that one sees important
space-time properties like general covariance.
An approach that we have been
pursuing is an extension of the renormalization group approach -
the `proper - time' approach \cite{bspt}.  We showed in \cite{bsos,bsfc}
how one can construct off-shell vertices in this approach
by keeping a finite cutoff.  One of the constraints used there was
gauge invariance.  In fact we observed an interesting fact - namely
that requiring electromagnetic gauge invariance in the presence
of a finite cutoff requires the presence of massive fields as
backgrounds.  Another constraint one should impose of course is that
these off-shell vertices be ``correct'' in the sense that they
can be used (with appropriate Feynman rules) to construct higher
N-point scattering amplitudes.

In this paper we would like to report on some progress in this second
aspect.
We show how one can insert delta functionals to factorize correlation
functions.  The delta functionals are conveniently represented as
loop variables of the kind introduced in \cite{bslv}.  Thus a four
tachyon scattering amplitude can be written as a sum of
terms of the form: vertex $\times$ propagator $\times$ vertex.
One can also extract
vertices where one of the particles is off-shell.
The idea of factorization is very old \cite{sc,l2},
however, our implementation is different.  Although in this paper
we will only discuss the case of four tachyon scattering, the
method is easily generalized to more complicated situations.
Of course,
off-shell vertices in open string theory have been discussed in the
context of string field theory \cite{sz,bp,w2,lpp}.
These vertices have nice geometric interpretations and they also reproduce
scattering amplitudes.  Nevertheless they do not have the simplicity
of the first quantized Polyakov formalism when it comes to doing
calculations. Our aim is to find something that is
 simpler to work with.  It is therefore probably worth exploring
objects that are directly obtained from scattering amplitudes, rather
than the other way around.

        In section II we discuss the harmonic oscillator to
illustrate the basic ideas.  In section III we discuss the string and
derive the main result.  In section IV we give some simple examples.
We conclude in section V with some comments and open questions.
\newpage
\section{The Harmonic Oscillator}
We want to understand the physical significance of certain
kinds of correlators, specifically ones that involve delta functions.
We will develop the basic idea by considering the example of a harmonic
oscillator.

The Lagrangian for a harmonic oscillator is
\be              \label{I}
{\cal L}= \frac{1}{2}(\dot{X}^{2} - \omega ^{2} X^{2})
\ee
where we have set the mass to unity.  Define, in the usual way,
the creation and annihilation operators:($p = \dot{X}$)
\be
a=\frac{ \sqrt{\omega}X + i \frac{p}{\sqrt{\omega}}}{\sqrt{2}}
\ee
\[
a^{+} = \frac{ \sqrt{\omega}X - i \frac{p}{\sqrt{\omega}}}{\sqrt{2}}
\]
\[
[a,a^{+}]=1  \,;\, \, [X,p]=i
\]
The Hamiltonian is
\be
H=\frac{1}{2}\omega (aa^{+} + a^{+}a)
\ee
and the ground state $\mid 0 >$, defined by $a \mid 0> =0$ is given
by the wave function:
\be
\Psi _{0} (X) = <X \mid 0> = (\frac{\omega}{\pi})^{\frac{1}{4}}
e^{-\frac{1}{2} \omega X ^{2}}
\ee

Now consider the following `one-point' function:
\be
O_{1} \equiv \int {\cal D} X \delta (X(t)-X^{I}) e^{-\frac{i}{\hbar}S}
\ee
where
$S = \int {\cal L}dt $ as in (\ref{I}).

In operator language we can represent it as (in the Schrodinger picture)
\be                         \label{VI}
O_{1}= <+\infty ,0 \mid U(+\infty ,t) \delta(X_{S}(t) - X^{I})
 U(t,-\infty )
\mid 0, -\infty >_{S}
\ee
(where the subscript `$S$'stands for ``Schrodinger'' and indicates
that $X$ is not a {\em function} of $t$).  We can also represent it
as
\be    \label{VII}
O_{1}= \int \frac{dk}{2 \pi} <+\infty ,0\mid U(+\infty ,t)
e^{ik(X_{S}(t) - X^{I})}U(t,-\infty)\mid 0,-\infty >_{S}
\ee
\be   \label{VIII}
= \int \frac{dk}{2 \pi} <0\mid e^{ikX_{H}(t)} \mid 0> _{H} e^{-ikX^{I}}
\ee
 where the subscript `$H$' stands for ``Heisenberg''.
 Note that $\mid 0>_{H} = \mid 0,0> _{S}$ (i.e. the vacua
 in the two pictures coincide at $t=0$)
 and
\be     \label{IX}
 X_{H}(t) = \frac{1}{\sqrt {2\omega}}(ae^{-i\omega t}+a^{+}e^{i\omega t})
\ee
 Due to the delta function in (\ref{VI}) we can rewrite $O_{1}$
 as
\be \label{X}
O_{1}=<+\infty ,0 \mid U(+\infty ,t)\mid X^{I},t><t,X^{I}\mid
U(t, -\infty )\mid 0,-\infty >_{S}
\ee
\[
=e^{i\omega t/2} <0\mid X^{I}><X^{I}\mid 0>_{H}e^{-i\omega t/2}
\]
\[
=\sqrt{\frac{\omega}{\pi}}e^{-\omega (X^{I})^{2}}
\]

 We can confirm this explicitly by calculating (\ref{VIII})
using
\be   \label{XI}
e^{ikX(t)}=:e^{ikX(t)}:e^{-k^{2}/4\omega}
\ee
and
\be  \label{XII}
<0\mid :e^{ikX(t)}:\mid 0>=1
\ee
Thus $O_{1}$ is the square of the ground state wave function.
Let us turn to the
 correlator of two delta functions:
Thus consider the following two objects:
\be          \label{XIII}
I=\int _{X(0)=X_{i}}^{X(T)=X_{f}}{\cal D}X(t) e^{-\frac{i}{\hbar} S}
\ee
namely the Feynman path integral for a harmonic oscillator
starting at $X_{i}$ at time $0$ and ending at $X_{f}$ at time $T$,
and:
\be    \label{XIV}
II=\int {\cal D}X(t) \delta (X(T)-X_{f}) \delta (X(0) - X_{i})
e^{-\frac{i}{\hbar} S}
\ee
the correlator of two delta functions, which also measures,
roughly speaking, the amplitude to go from one point to another.
Using arguments similar to that used for the one point function
one can show that the
the two are related as follows:
\be         \label{XV}
II = (\frac{\omega}{\pi})^{1/2}e^{i\omega T/2}e^{-\frac{\omega}{2}
(X_{i}^{2}+X_{f}^{2})} I
\ee
This result is easily verified by explicit calculation:
I is a quantity that can be found in textbooks.  II is easily calculated
using the vertex operator representation of the delta function and the
correlation between vertex operators:
\[
<0\mid :e^{ikX(t_{2})}: :e^{ipX(t_{1})}: \mid 0>
\]
\[
e^{-kp<X(t_{2})X(t_{1})>}
\]
\[
e^{-kp\frac{e^{-i\omega (t_{2}-t_{1})}}{2\omega}}
\]

This result is useful because objects like I are what we
have in string field theory -``propagator''- whereas objects like
II  are what are easy to calculate in conformal field theory.
  We now proceed to apply these ideas to strings.

\newpage
\section{Factorizing String Amplitudes}
\setcounter{equation}{0}

Consider the following four-point function:
\be       \label{1}
<0\mid :V_{1}(z)::V_{2}(u)::V_{3}(v)::V_{4}(w):\mid 0>
\ee
Here $:V_{i}:$ are normal ordered vertex operators used in string
scattering amplitude calculation.
The usual Veneziano amplitude is obtained by fixing $z=\infty , u=1,
w=0$, multiplying by $z^2$ and integrating over $v$ from $0$ to $1$.
In the proper-time equation \cite{bspt}, one would integrate
$u$ and $v$.  In either case, one has to first factorize (\ref{1}).
 We begin by inserting two delta functionals of the form:
 \be     \label{2}
 \prod_{n=0}^{\infty} \delta(X_{n}(\tau _{1})-X_{n}^{I})
 \prod_{n=0}^{\infty} \delta(X_{n}(\tau _{2})-X_{n}^{II})
 \ee
 Here $X_{n}$ is the harmonic oscillator coordinates of the $n$th
 mode of the string.  To be definite we consider an open string
 with a mode expansion:
 \[
 X(t,\sigma )=\frac{X_{0}(t)}{\sqrt{2}} + \sum _{n>0}
 X_{n}(t)cos n\sigma
\]
 \be          \label{3}
 =\frac{x_{0}+pt}{\sqrt{2}} + \sum _{n>0}
 \frac{1}{\sqrt{2n\omega}}(a_{n}^{+}e^{in\omega t} + a_{n}e^{-in\omega t})
 cos n\sigma
 \ee
 We can rewrite this as
 \be   \label{4}
 X(z,\tilde{z})=\frac{x_{0}}{\sqrt{2}} -\frac{iplnz +ipln\tilde{z}}
 {2\sqrt{2}} +
 \frac{1}{2} \sum _{n>0}\frac{1}{\sqrt{2n\omega}}(a_{n}^{+}
 z^{n} + a_{n}z^{-n})+
 \ee
 \[
 \frac{1}{2} \sum _{n>0}\frac{1}{\sqrt{2n\omega}}(a_{n}^{+}
 \tilde{z}^{n} + a_{n}\tilde{z}^{-n})
 \]
 where
 $z= e^{i(\omega t + \sigma )}$ and $\tilde{z} =
  e^{i (\omega t - \sigma )}$
 If we analytically continue $t \rightarrow -i\tau $ to the
 Euclidean world sheet, then $\tilde{z} = \bar{z}$.
 We can then write
 \be             \label{5}
 X(z,\bar{z})=\frac{1}{2}(X(z)+X(\bar{z}))
 \ee
 where
 \be             \label{6}
 X(z)=\frac{x_{0}}{\sqrt{2}} -\frac{iplnz}
 {\sqrt{2}} +
  \sum _{n>0}\frac{1}{\sqrt{2n\omega}}(a_{n}^{+}
 z^{n} + a_{n}z^{-n})
 \ee
We will work with just $X(z)$, with the understanding that
 for the open string one has to add the complex conjugate.  Of course,
 most of the time this has no effect because the vertex operators
 are at the boundary of the world sheet where $\sigma =0,\pi $ and
 $z = \bar{z}$.
After inserting the delta functionals we get:
\[
\int [dX_{n}^{I}][dX_{n}^{II}]
<0\mid :V_{1}(z)::V_{2}(u):
 \prod_{n=0}^{\infty} \delta(X_{n}(\tau _{1})-X_{n}^{I})
 \]
 \be    \label{7}
 \prod_{n=0}^{\infty} \delta(X_{n}(\tau _{2})-X_{n}^{II})
:V_{3}(v)::V_{4}(w):\mid 0>
\ee
\[
= \int [dX_{n}^{I}][dX_{n}^{II}] \underbrace{
<0\mid :V_{1}(z)::V_{2}(u):\mid [X_{n}^{I}],\tau _{1}>}_{I}
\]
\[
\underbrace{
<[X_{n}^{I}],\tau _{1}
\mid U(\tau _{1},\tau _{2})\mid [X_{n}^{II}],\tau _{2}>}_{II}
\]
\be        \label{8}
\underbrace{
< [X_{n}^{II}],\tau _{2}
\mid :V_{3}(v)::V_{4}(w):\mid 0>}_{III}
\ee
 Consider I:  It can be rewritten as:
 \be    \label{9}
 <0\mid :V_{1}(z)::V_{2}(u):\mid [X_{n}^{I}],\tau _{1}>
 \underbrace{<[X_{n}^{I}],\tau _{1}
 \mid U(\tau _{1},0)\mid 0>}_{{\cal N}_{1}}
 {\cal N}_{1}^{-1}
 \ee
 where
 \be    \label{10}
 {\cal N}_{1}=\prod _{n} (\frac{n\omega}{\pi})^{1/4}
 e^{-\frac{n\omega (X_{n}^{I})^{2}}{2}}e^{-\frac{in\omega \tau _{1}}{2}}
 \ee
 Thus,
 \be    \label{11}
I=<0\mid :V_{1}(z)::V_{2}(u):
 \prod_{n=0}^{\infty} \delta(X_{n}(\tau _{1})-X_{n}^{I})\mid 0>
  {\cal N}_{1}
 ^{-1}
 \ee
 \be    \label{12}
= \prod_{n=0}^{\infty}\int \frac{dk_{n}}{2\pi}
 <0\mid :V_{1}(z)::V_{2}(u):
  e^{ik_{n}X_{n}(\tau _{1})}\mid 0>
  e^{-ik_{n}X_{n}^{I}} {\cal N}_{1}^{-1}
 \ee
 \be    \label{13}
 =\prod_{n=0}^{\infty}\int \frac{dk_{n}}{2\pi}
 <0\mid :V_{1}(z)::V_{2}(u):
 :e^{ik_{n}X_{n}(\tau _{1})}:\mid 0>
  e^{-ik_{n}X_{n}^{I}} e^{-\sum_{m>0}\frac{k_{m}^{2}}{4 \omega _{m}}}
  {\cal N}_{1}^{-1}
 \ee
 This is the final form.  One can also rewrite
 \be    \label{14}
 e^{ik_{n}X_{n}(\tau _{1})}  = e^{i\int dt k(t) X(at)}
 \ee
 where $a= e^{i\omega \tau _{1}}$, $k(t) =\Sigma _{n} k_{n} t^{-n-1}$
 and $k_{n}=k_{-n}$.
 In this form one recognizes the loop variable introduced in \cite{bslv}.
 The normal ordering of these variables was described in \cite{bsvg},
 and for the case where $k_{n}=k_{-n}$ it can be seen to be the same
 as in (\ref{13}).  One can also write (\ref{14}) as
 \be    \label{15}
 e^{i\sum_{n}a^{n}k_{n}\frac{\partial ^{n} X(0)}{n!}}
 \ee
 However this is only valid provided there is no singularity as
 $a\rightarrow 0$, which will be the case if there is no other
 vertex operator at $z=0$.  Performing similar manipulations on the
 other two factors one gets:
 \[
\int \prod _{n}dX{n}^{I} dX{n}^{II}  \frac{dk_{n}}{2\pi}
 \frac{dp_{n}}{2\pi}  \frac{dq_{n}}{2\pi}  \frac{dl_{n}}{2\pi}
  e^{-ik_{n}X_{n}^{I}-ip_{n}X_{n}^{I}
  -il_{n}X_{n}^{II}-iq_{n}X_{n}^{II}}
  e^{-\sum_{m>0}(\frac{k_{m}^{2}}{4 \omega _{m}}+
 \frac{p_{m}^{2}}{4 \omega _{m}} +
  \frac{l_{m}^{2}}{4 \omega _{m}} +
  \frac{q_{m}^{2}}{4 \omega _{m}})}
\]
\[
  \underbrace
  {<0\mid :V_{1}(z)::V_{2}(u):e^{ik_{n}X_{n}(\tau _{1})}\mid 0>}_{\cal V}
  \underbrace{
< 0\mid :e^{ip_{n}X_{n}(\tau _{1})}:: e^{iq_{n}X_{n}(\tau _{2})}:\mid 0>}
_{\cal P}
 \]
 \be    \label{16}
 \underbrace {
 <0\mid :e^{il_{n}X_{n}(\tau _{2})}V_{3}(v)::V_{4}(w):
  \mid 0>}_{\cal W}
 {\cal N}_{1}^{-1} {\cal N}_{1}^{*-1}{\cal M}_{1}^{-1}{\cal M}_{1}^{*-1}
 \ee
 Here, $\cal M$ is given by (\ref{10}) with $X^{I}, \tau _{1}$
 replaced by $X^{II}, \tau _{2}$ and $*$ refers to complex conjugation.
 One can check by explicit calculation, for particular choices of the
  $V_{i}$
 that (\ref{16}) reproduces the
 right result.  While this is guaranteed by construction,
  it is illuminating
 to see how it happens.
 In fact it gives a proof of the formula for the operator product
 of two vertex operators \cite{poly}.
 We will not do it here, though.  Modulo
 factors required to amputate external legs (when they are off-shell)
 (\ref{16}) can be recognized to be in the form vertex $\times$ propagator
 $\times$
 vertex as we will see below.  (The propagator emerges after integrating
 over $v$).  The normal ordering of the zero modes requires some care.
 Vertex operators in string theory are normal ordered - which for the
 zero modes usually means that $p$ is to the right of $x$.
 \footnote{Note, however, that
 in (\ref{16}), the factors $e^{ik_{0}x_{0}(\tau_{1})}$ obtained
 as a Fourier transform of the delta functions, are {\em not} normal
 ordered (remember that we are only talking about the zero modes).}
 In fact the two three point vertices in (\ref{16}) are asymmetric
 with respect to the zero modes precisely because of this ordering.
 This asymmetry disappears when the external particles are on shell
 even if the internal one is not.
 It is presumably possible to adopt ordering conventions that
treat both vertices in a symmetric manner.
 \newpage
 \section{Examples}
 \setcounter{equation}{0}
 We will consider the case where the external particles are on-shell
 tachyons
 so that we are calculating an S-matrix element. Thus,
 $V_{1}=e^{i\bar{k}X}$, $V_{2}=e^{i\bar{p}X}$,
 $V_{3}=e^{i\bar{q}X}$ ,$V_{4}=e^{i\bar{l}X}$.  This will give us vertices
 with one off-shell and two on-shell particles.  We will choose
 $e^{i\omega \tau_{1}} =u$ and $e^{i\omega \tau_{2}}=v$.  This will
 remove as much of the $v$-dependence as possible from the vertex.
 In that case we get
 \be    \label{17}
{\cal V}=(1-\frac{u}{z})^{\frac{\bar{p}\bar{q}}{2\omega}}(\frac{z}{u})^{-
\frac{\bar{k}^{2}}{2\omega}}u^{-\frac{\bar{p}^{2}+\bar{k}^{2}}{4\omega}}
e^{-(\frac{k_{n}\bar{k}(\frac{u}{z})^{n} + k_{n}\bar{p}}{2n\omega})}
 \ee
 In the same way $\cal W$ becomes
 \be    \label{18}
{\cal W}=(1-\frac{w}{v})^{\frac{\bar{q}\bar{l}}{2\omega}}
v^{-\frac{\bar{l}^{2}+\bar{q}^{2}}{4\omega}}
e^{-(\frac{l_{n}\bar{l}(\frac{w}{v})^{n} + l_{n}\bar{q}}{2n\omega})}
 \ee
 The vertices do not look quite the same.  However when the external
 particles are on shell (but not the internal one) one can choose
 $z= \infty, u=1 \,\, and\,\, w=0$.  This simplifies the vertex
 enormously.  Note that we have to multiply $\cal V$ by $z^{2}$.
  Furthermore
 we can multiply $\cal W$ by $v^{2}$ and compensate by dividing
 $\cal P$ by
 $v^{2}$.  At the end we are left with a remarkably simple form for
 the vertex:
 \be    \label{19}
\prod _{n}e^{-\frac{k_{n}\bar{p}}{2n\omega}}
 \ee
The two point function $\cal P$ (divided by $v^{2}$) takes the form
\be     \label{20}
\prod _{n}
\frac{1}{v^{2}}e^{-\frac{p_{n}q_{n}}{2n\omega}(\frac{v}{u})^{n}}(\frac
{u}{v})^{-\frac{p_{0}^{2}}{2\omega}}
\ee
One can put the factors $\cal V$,$\cal W$,$\cal P$ together and do the
integrals in
(\ref{16}).  The integrals are fairly straightforward.  The noteworthy
feature being that the the normal ordering factors cancel the wave
function
factors. One ends up with delta functions (in $p_{n} ,q_{n}$)
and their derivatives multiplied by polynomials in $p_{n}, q_{n}$ from
expanding the exponent in (\ref{20}).
\subsection{Tachyon-tachyon-tachyon}
In the case where the intermediate state is a tachyon, on sets all
$k_{n} =0 = l_{n}$.  This gives  $1$ for the vertex, and the
propagator is just
\be     \label{21}
\int _{0}^{1} dv v^{\frac{p_{0}^{2}}{4\omega}-2} =
\frac{1}{\frac{p_{0}^{2}}{4\omega}-1}
\ee
(where $p_{0}=\bar{p}+\bar{k}$)as expected.
\subsection{Tachyon-tachyon-vector}
We choose one factor of $k_{1},l_{1}$ from the vertex.  This gives:
\[
\int \frac{dp_{1}dq_{1}}{4\pi ^{2}}
\bar{p}^{\mu}[\frac{d}{dp_{1}^{\mu}}\frac{d}{dq_{1}^{\nu}}
\delta(p_{1})\delta (q_{1})] p_{1}q_{1}\frac{1}{\frac{p_{0}^{2}}{4\omega}}
\bar{q}^{\nu}
\]
\be     \label{22}
= \frac{\bar{p}\bar{q}}{4\pi^{2}\frac{(\bar{p}+\bar{k})^{2}}{4\omega}}
\ee
as expected.
Note that the vector-tachyon-tachyon vertex has the asymmetric form
$\bar{p}.k_{1}$ and not $(\bar{p} - \bar{k}).k_{1}$. This is not
surprising given the asymmetric form of (\ref{16}).  However, by
construction, it is guaranteed to reproduce the Veneziano amplitude.

Proceeding as above it is easy to get interaction vertices for other
particles, both internal and external.
          Note that our factorization is in a particular channel.
Of course in string theory it is sufficient to sum over only one
channel, so this is consistent.
\newpage
\section{Conclusions}

We have constructed a class of vertices by factorizing tree amplitudes
and by the
nature of the construction they are guaranteed to give the right answer
for scattering amplitudes.
  The vertices have one
particle off-shell and two on shell.  This is because we started with
an on-shell S-matrix element. By considering scattering amplitudes
with more than four particles one can construct vertices where two
particles are off-shell.  The vertices obtained here, are not very
symmetric looking, nevertheless they have the advantage of being
very simple and do not require much effort to work out.

There are many open questions that arise.  The most important is to
extend it to the case where all the particles are off-shell.
One would like to compare this with the finite cutoff vertices\cite{bsfc}.
Perhaps if the calculation is done on a disc rather than the half-plane
the results will look more symmetrical.  Finally, it would be interesting
to see if it is possible to
implement gauge invariance along the lines of \cite{bslv}.

\end{document}